\documentclass[conference]{IEEEtran}
\IEEEoverridecommandlockouts
\usepackage{xcolor}

\usepackage{cite}
\usepackage{amsmath,amssymb,amsfonts}
\usepackage{algorithm}
\usepackage{algpseudocode}
\usepackage{graphicx}
\usepackage{textcomp}
\usepackage{xcolor}
\usepackage{soul}
\usepackage{enumitem}
\usepackage{pifont}
\usepackage{stfloats}
\usepackage{url}

\usepackage{flushend}

\usepackage{booktabs}
\usepackage{multirow}
\usepackage{tabularx}
\usepackage{diagbox}
\usepackage{colortbl}

\usepackage{makecell}

\usepackage[caption=false]{subfig}
\usepackage[font=footnotesize,labelfont=bf]{caption}

\def\BibTeX{{\rm B\kern-.05em{\sc i\kern-.025em b}\kern-.08em
    T\kern-.1667em\lower.7ex\hbox{E}\kern-.125emX}}

\usepackage[absolute]{textpos}

\begin{document}


\begin{textblock}{5}(11.8,0.55)
(Special Session)
\end{textblock}

\begin{textblock}{14}(5.7,0.75)
This paper will be presented at IEEE VLSI Test Symposium (VTS) 2026.
\end{textblock}


\title{Secure eFPGA-Enabled Edge LLM Inference: Architectural and Hardware Countermeasures}

\author{\IEEEauthorblockN{Voktho Das, M Zafir Sadik Khan, Jafar Vafaei, Kimia Azar, Hadi Kamali}
\IEEEauthorblockA{\textit{Department of Electrical and Computer Engineering (ECE), University of Central Florida, Orlando, FL 32816, USA} \\
\{voktho.das, mzafirsadik.khan, jafar.vafaei, azar, kamali\}@ucf.edu}
}

\maketitle

\begin{abstract}
Edge deployment of transformer-based models increasingly relies on ASIC accelerators due to their high performance and energy efficiency, achieved through optimized dataflows, specialized architectures, low-bitwidth computation, and efficient memory hierarchies. However, these advantages come with significant security vulnerabilities. ASIC-based DNN accelerators are susceptible to side-channel attacks (e.g., power, electromagnetic, and timing analysis) and fault injection attacks (e.g., voltage manipulation, clock glitches, and memory perturbations), which can lead to model extraction or compromised inference integrity. Furthermore, threats introduced during design and fabrication, such as hardware Trojans or untrusted third-party IPs, further expand the attack surface. To address these challenges, we explore a hybrid ASIC+eFPGA architecture that combines the efficiency of ASICs with the flexibility of reconfigurable logic. The integrated eFPGA enables security-oriented mechanisms such as adaptive runtime monitoring, side-channel mitigation and post-deployment patching. By leveraging these capabilities, the proposed approach enhances system resilience against both runtime and supply-chain attacks, while preserving the performance benefits of ASIC-based transformer inference.
\end{abstract}

\begin{IEEEkeywords}
Transformers, eFPGA Acceleration, Security.
\end{IEEEkeywords}

\section{Introduction}

Hardware accelerators drive modern AI, with GPUs, FPGAs, and custom ASICs widely deployed for high-throughput, energy-efficient execution of neural networks \cite{mishra2023artificial, dhilleswararao2022efficient}. Among these, ASICs are increasingly favored in both cloud and edge settings for real-time inference due to their superior power-performance characteristics \cite{Parashar2019Timeloop}. However, deploying DNNs on ASICs introduces notable security challenges \cite{potluri2024sok_eprint913}. Prior work shows that adversaries can extract model architectures via microarchitectural side-channels (e.g., DRAM access patterns) \cite{yan2023defense}, infer models from timing-based leakage in inference latency \cite{Won2021TimeToLeak}, and even perform remote side-channel attacks by capturing power traces and reconstructing model details using learning-based techniques \cite{yan2023mercury}.

Meanwhile, LLMs are rapidly scaling to hundreds of trillions of parameters \cite{fedus2022switch}, placing heavy demands on compute and memory and making hardware acceleration essential. Recent work shows that advances in LLM capabilities are closely tied to specialized hardware \cite{guo2025survey}. As a result, diverse accelerators, including GPUs, FPGAs, ASICs, and processing-in-memory (PIM) systems, have been explored to efficiently execute transformer workloads \cite{li2024large}. While FPGAs enable flexible LLM accelerators, ASIC-based transformer hardware can achieve higher throughput and energy efficiency through domain-specific datapaths and operator optimizations, particularly for high-performance LLM inference.

While ASIC-based LLM inference enables real-time deployment at cloud and edge scale, a key question remains: \textit{Can ASIC efficiency be combined with the adaptability of reconfigurable logic to improve hardware security?} Specifically, can an ASIC LLM accelerator with a small embedded FPGA (eFPGA) fabric mitigate attacks affecting conventional ASIC-based DNN deployments? This paper investigates the security benefits of eFPGA-enhanced LLM accelerators by examining common hardware attacks targeting ASIC-based DNN accelerators and analyzing how embedded FPGA integration can be used to mitigate these threats. We then show how eFPGA can enable architectural countermeasures, including selective logic redaction, runtime monitoring, adaptive fault response, and post-deployment patching. The proposed hybrid design aims to retain ASIC-class performance for the performance-critical datapath while introducing a constrained and auditable reconfigurable security module. Finally, we discuss the advantages and limitations of these approaches and highlight open challenges for secure LLM accelerator design.

\begin{figure}[b]
\centering
\vspace{-5pt}
\includegraphics[width=\linewidth]{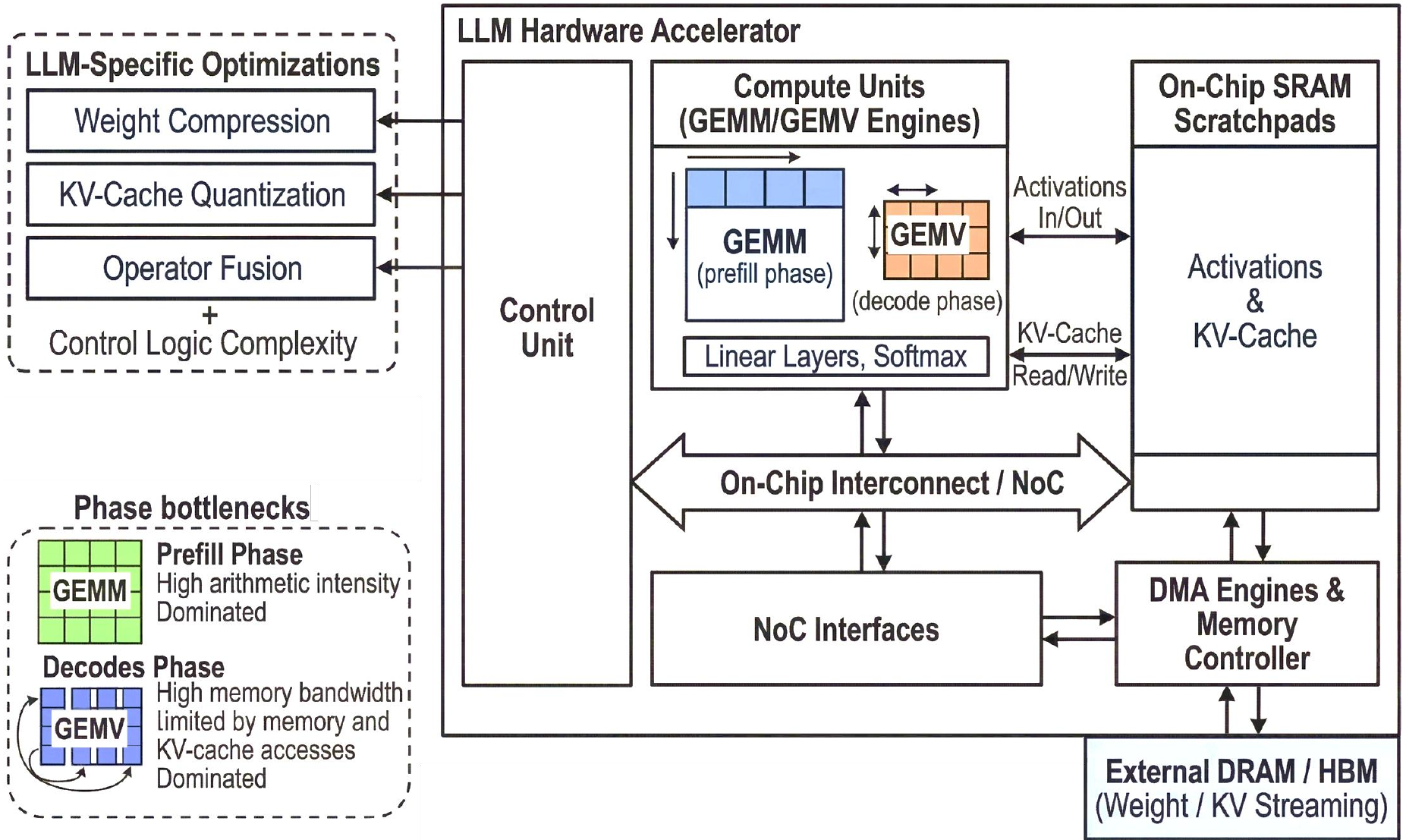}
\caption{Comprehensive Architecture of Decoder-Only LLM Accelerators.}
\label{fig:llm_hw_view}
\end{figure}
 
\section{Prior Art and Threat Model}

\subsection{Transformer/LLM Accelerator Anatomy}

Modern decoder-only LLM inference typically consists of two phases with different bottlenecks: prefill, which exhibits high arithmetic intensity, dominated by general matrix multiplication (GEMM) operations, and decode, which is often limited by memory bandwidth and KV-cache accesses, dominated by general matrix-vector multiplication (GEMV) operations \cite{li2024large, he2025waferllm}. Across both phases, the accelerator repeatedly executes linear layers (e.g., QKV projections and MLPs), attention score computation and softmax operations, as well as frequent KV-cache reads/writes, which is increasingly treated as a hardware–software co-design problem \cite{he2025waferllm, kim2025oaken}. As shown in Fig.~\ref{fig:llm_hw_view}, transformer inference hardware, either FPGA or ASIC, typically includes: (i) compute units, (ii) on-chip SRAM scratchpads for activations and KV-cache, (iii) DMA engines and memory controllers for weight/KV streaming, (iv) an on-chip interconnect or NoC, and (v) a control unit for scheduling, quantization, sparsity, and dataflow management \cite{li2024large, koilia2024hardware}. In this generic architecture, LLM-specific optimizations, e.g., weight compression, KV-cache quantization, and operator fusion, further increase the complexity of the control unit. As a result, small perturbations or faults in control logic can propagate through the execution pipeline, leading to significant semantic deviations in generated outputs.

\begin{table*}[bp]
\centering
\scriptsize
\setlength{\tabcolsep}{3pt}
\vspace{-10pt}
\caption{Attack vectors vs.\ Affected LLM Accelerator Components and eFPGA+ASIC Mitigation. Goals: C=confidentiality, I=integrity, A=availability.}
\begin{tabular}{@{} p{61pt} p{16pt} p{113pt} p{120pt} p{180pt}@{}}
\toprule
\textbf{Attack vector} & \textbf{Goal} & \textbf{Affected LLM components} & \textbf{Enablers / observations} & \textbf{Recommended eFPGA mitigations} \\
\cmidrule[\heavyrulewidth](r){1-1} \cmidrule[\heavyrulewidth](r){2-2} \cmidrule[\heavyrulewidth](r){3-3} \cmidrule[\heavyrulewidth](r){4-4} \cmidrule[\heavyrulewidth](r){5-5} 
Voltage droop / timing faults \cite{rakin2021deep, boutros2020neighbors} & I/A &
Weight DMA, on-chip buffers, MAC array timing, control FSM &
Co-tenant power-plundering; clock gating transients can evade bitstream checks \cite{boutros2020neighbors} &
eFPGA runtime monitors; reconfigurable guardbands; fault-aware microcode patching \\
\cmidrule(r){1-1} \cmidrule(r){2-2} \cmidrule(r){3-3} \cmidrule(r){4-4} \cmidrule(r){5-5} 
Remote side-channels \cite{tian2023practical} & C &
Inputs, activations, layer schedule, memory bursts &
TDC/RO sensors infer activity remotely; recognizable runtime patterns &
eFPGA equalization,
masking, and perturbation; runtime monitors \\
\cmidrule(r){1-1} \cmidrule(r){2-2} \cmidrule(r){3-3} \cmidrule(r){4-4} \cmidrule(r){5-5} 
Physical side-channels  \cite{chaudhuri2025energon} & C &
MAC/systolic arrays, on-chip SRAM, off-chip memory accesses &
CPA reveals weights; GPU power channels leak transformer structure \cite{chaudhuri2025energon} &
eFPGA equalization,
masking, and perturbation\\
\cmidrule(r){1-1} \cmidrule(r){2-2} \cmidrule(r){3-3} \cmidrule(r){4-4} \cmidrule(r){5-5} 
\textmu architectural leak in inference \cite{gao2025know} & C &
Token embedding tables; KV-cache dependent control flow &
Token value/position leakage via cache access patterns &
eFPGA equalization, masking, and perturbation; post-deployment patching \\
\cmidrule(r){1-1} \cmidrule(r){2-2} \cmidrule(r){3-3} \cmidrule(r){4-4} \cmidrule(r){5-5} 
Single-bit flips in quantization \cite{gongye2024one} & I/A &
Quantization scaling factors; code regions; control metadata &
Scaling factors become SPOFs; code regions can be more fragile than data \cite{gongye2024one} &
eFPGA TEE; selective redundancy/ECC for control/metadata; post-deployment patching \\
\cmidrule(r){1-1} \cmidrule(r){2-2} \cmidrule(r){3-3} \cmidrule(r){4-4} \cmidrule(r){5-5} 
Hardware Trojans (3PIP) \cite{mukherjee2022novel} & I/C/A &
Activation paths, interconnect, control logic, triggers &
Trigger-based payloads corrupt activations; difficult post-silicon assurance &
selective IP redaction; eFPGA TEE; runtime monitors  \\
\cmidrule(r){1-1} \cmidrule(r){2-2} \cmidrule(r){3-3} \cmidrule(r){4-4} \cmidrule(r){5-5} 
IP piracy / reverse engineering \cite{mohan2021hardware, kamali2023shell, das2025soar} & C &
Unique control logic, proprietary ops (packing, sparsity), security policy logic &
Untrusted fab/integrator can recover netlist or insert clones &
eFPGA IP redaction; device-bound bitstream encryption (PUF)\\ 
\bottomrule
\end{tabular}
\label{tab:attack_vs_counter}
\end{table*}

\subsection{Threat Model and Priorities}

We consider an adversary capable of exploiting hardware and microarchitectural vulnerabilities in ASIC-based DNN accelerators across the design, fabrication, and runtime phases. The attacker may operate remotely in multi-tenant environments, leverage physical access for side-channel and fault injection attacks, or act within the supply chain through untrusted IPs or fabrication processes. This power enables attacks such as side-channel leakage, fault-induced errors, and hardware Trojan insertion targeting model weights, intermediate data, and control logic. We assume the base ASIC may not be fully trusted, while the integrated eFPGA fabric is trusted (e.g., through secure boot) and used to monitor, detect, and mitigate such threats at design and runtime.

In LLM accelerators, key confidentiality assets include (i) model weights and quantization metadata, (ii) the KV cache containing user prompts and contextual information, and (iii) control-plane artifacts such as microcode and scheduling heuristics. Beyond confidentiality, the architecture must also ensure integrity and availability by preventing (iv) output corruption, (v) targeted output manipulation or denial-of-service, and (vi) stealthy long-term degradation caused by single-point failures or hardware Trojan triggers.

Based on these priorities and the defined threat model, we analyze how possible attack scenarios impact LLM accelerators. Table \ref{tab:attack_vs_counter} summarizes the attack vectors, the potentially targeted components of the LLM, and the corresponding mitigations enabled by the proposed ASIC+eFPGA architecture.

\section{Why eFPGA+ASIC Can Help?}

\subsection{eFPGA Fabrics Inside an ASIC LLM accelerator}

Prior work shows that ASIC-based LLM accelerators achieve high throughput and energy efficiency via domain-specific specialization \cite{zhu2020powerscout}. However, ASIC-only designs remain vulnerable to diverse attack vectors, including hardware-level threats (e.g., Trojans, side-channel attacks, counterfeiting, reverse engineering) and architectural weaknesses such as insecure memory hierarchies, shared-resource leakage, and limited isolation \cite{gao2025know, chaudhuri2025energon}. This raises a key challenge: \textit{how to introduce security agility into ASIC accelerators using eFPGA to mitigate their broad attack surface.}

We argue that eFPGA offers a promising solution as a reconfigurable security and auxiliary module. The embedded fabric is (i) compact and tailored for specific functions, and (ii) configured exclusively through a trusted boot and update chain to prevent unauthorized modifications.

\subsection{Mitigation concepts enabled by eFPGA}

The reconfigurability of eFPGA fabrics enables a wide range of hardware and architectural countermeasures against attacks. We outline five key mitigation strategies that leverage this flexibility to enhance the security of LLM accelerators. 

\noindent \textbf{\textit{\underline{S1. Selective IP redaction of Critical Logic.}}} IP redaction using eFPGA fabric is a conventional approach to mitigate threats arising during both manufacturing and deployment stages, including reverse engineering, overproduction, and counterfeiting \cite{kamali2022advances}. In this approach, security-critical logic blocks of the original design are partitioned and replaced with an eFPGA macro, while the corresponding configuration bitstream is withheld until trusted provisioning \cite{sami2024advancing}. In LLM accelerators, these security-critical sections may include complex routing logic or specialized units such as sparsity handling and quantization modules \cite{kamali2023shell, das2026nuredact, Gubbi2023SecuringAI}. By reducing the visibility of these components to the fabrication facility or system integrator, this method enhances design confidentiality and strengthens protection against unauthorized access, cloning, or replication. Additionally, although eFPGA-based obfuscation primarily protects the design, it also provides resilience against hardware Trojans by reducing observability and making insertion points harder to identify. It further enhances resistance to side-channel attacks (e.g., power, timing, EM) by obscuring internal behavior, making leakage harder to exploit. 

\begin{figure*}[t]
\centering
\includegraphics[width=\linewidth]{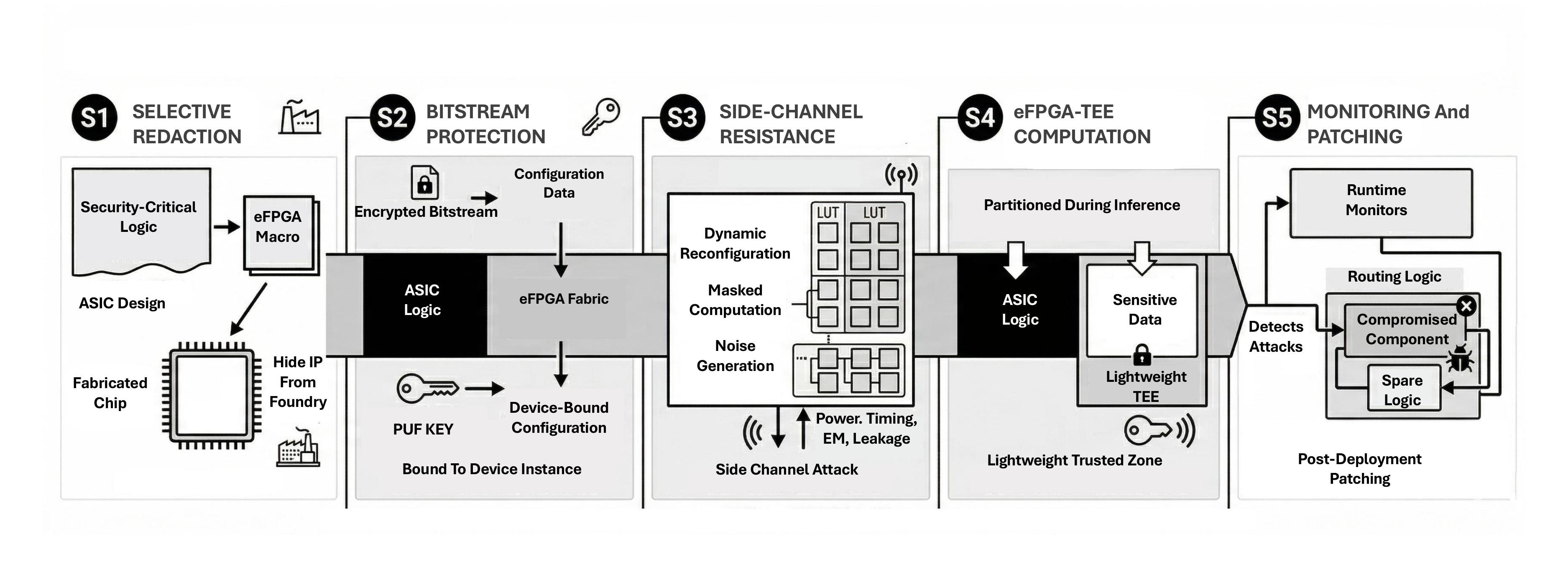}
\vspace{-28pt}
\caption{End-to-end Secure eFPGA-enabled ASIC Architecture illustrating a Defense Strategy across the Hardware Lifecycle.}
\vspace{-15pt}
\label{fig:hybrid_solution}
\end{figure*}

\noindent \textbf{\textit{\underline{S2. Encrypted/Device-bound eFPGA Configuration.}}} The eFPGA fabric functionality can be protected through encrypted and device-bound configuration using a device-unique key derived from a physically unclonable function (PUF), such as an SRAM PUF, instead of relying on static keys stored in non-volatile memory. As an example, commercial platforms, such as FlexLogix has shown that in this model, a root key is reconstructed at boot from the chip’s physical fingerprint and used to decrypt and authenticate the eFPGA configuration bitstream \cite{flexlogixpuf2022}. This mechanism prevents unauthorized redistribution of the configuration, as the bitstream becomes bound to a specific device instance. It also ensures the integrity of the configured eFPGA logic, which is particularly important when the fabric hosts security-critical functionality such as redacted control logic or runtime security monitors. PUF-based key derivation requires fuzzy extractors for stability across voltage, temperature, and aging, and provisioning during manufacturing tests. With an existing hardware root-of-trust (RoT), this overhead is modest relative to the security benefits. Furthermore, with modular integration and tiled fabrics, the reconfigurable region can be sized appropriately for security without significantly impacting the accelerator datapath.

\noindent \textbf{\textit{\underline{S3. Side-Channel Resistance via Reconfigurable Fabric.}}} A variety of techniques have been proposed to improve side-channel resistance (SCR) in deep learning accelerators \cite{potluri2024sok_eprint913,latibari2024transformers_security}, generally categorized into equalization, masking, and perturbation \cite{maji2022threshold}.

Equalization techniques reduce leakage by making side-channel signals independent of processed data. In ASICs, this is achieved using circuit-level countermeasures such as DPA-resistant logic and complementary operand computation \cite{Tiri2002Dynamic,DoulcierVerdier2011AES}. In eFPGA, similar effects can be realized through dynamic partial reconfiguration (DPR) and netlist-level diversity, where functionally identical but structurally different implementations are switched at runtime \cite{asghar2022benefits, Asghar2021Increasing}. This reconfigurability helps disrupt consistent leakage patterns. 

Masking techniques hide sensitive information by splitting operands into multiple randomized shares that are statistically independently and recombined at the output (e.g., arithmetic and Boolean masking) \cite{Dubey2020BoMaNet,Dubey2022Guarding,Dubey2022ModuloNET}. This approach integrates well with ASIC-eFPGA systems, where masking operations can be distributed across reconfigurable logic, reducing correlation between leakage and secret data.

Perturbation techniques introduce noise into a system to obscure side-channel leakage, making it significantly harder for an attacker to extract useful information from power, electromagnetic, or timing signals. eFPGA fabrics enable on-chip noise generators—such as toggled LUT chains or ring-oscillator-based engines—using unused resources\cite{yan2023defense}. These reduce the signal-to-noise ratio (SNR) of side-channel emissions and can be deployed independently of workloads.

Beyond these, architectural countermeasures include operation shuffling, which randomizes execution order to hinder trace alignment\cite{ganesan2023blackjack}, and constant-shape traffic generation for DMA or NoC to hide memory access patterns\cite{yan2023defense}.

\noindent \textbf{\textit{\underline{S4. Heterogeneous Computation on eFPGA TEEs.}}} Ensuring both confidentiality and integrity during DNN inference has driven extensive research on trusted execution environments (TEEs)\cite{Lee2023SecureLoop}. TEEs provide strong security guarantees by isolating sensitive computations and enforcing secure memory access, protecting model parameters and intermediate data from untrusted components. However, executing the full inference pipeline within a TEE incurs significant performance and energy overheads. To mitigate this, heterogeneous inference partitions computation into sensitive and non-sensitive components, where the sensitive portion runs inside the TEE while the rest is offloaded to a high-performance accelerator\cite{Bai2025Phantom,Nayan2025SecureInfer}. This selective protection improves the trade-off between security and efficiency by restricting the TEE boundary to critical operations. In embedded systems with eFPGA, this paradigm can be extended by using reconfigurable fabric as a lightweight TEE. Security-critical computations—such as those involving confidential weights or sensitive activations—can be mapped to the eFPGA, while throughput-critical kernels remain on the ASIC. This approach suits resource-constrained environments by enabling fine-grained security partitioning without full-system protection. Moreover, FPGA platforms efficiently support cryptographic primitives, enabling encryption, authentication, and integrity verification\cite{ahmed2025lightweight}. Overall, this architecture balances security, performance, and flexibility for emerging LLM inference workloads.

\noindent \textbf{\textit{\underline{S5. Runtime Monitors and Post-deployment Patching.}}} By placing reconfigurable logic near sensitive assets, eFPGA fabrics enable both runtime monitoring and post-deployment adaptation to detect and mitigate hardware attacks. Runtime monitors—such as timing, voltage, control-flow, and metadata integrity checks—can identify faults induced through mechanisms like power delivery network (PDN) manipulation or clock-gating transients, as well as corruption of critical data (e.g., quantization metadata) that can significantly impact inference accuracy \cite{rahman2023efficient, rahman2024road, boutros2020neighbors, rakin2021deep, gongye2024one}. This capability is particularly important for LLMs, where small numerical errors can propagate through autoregressive decoding and lead to large semantic deviations. In addition to detection, eFPGA fabrics provide spare logic for rerouting around compromised components (e.g., hardware Trojans or aging-induced faults) and enable post-deployment security patching without requiring full ASIC redesign \cite{das2025soar}. This supports long-lived edge deployments and evolving LLM workloads, where new vulnerabilities may emerge over time \cite{gao2025know, chaudhuri2025energon}.

Ensuring robust protection requires carefully calibrated detection thresholds to avoid false positives, as well as verifiable patch pipelines and formal constraints to prevent misuse of the reconfigurable fabric \cite{liu2022hardware}. Fig.~\ref{fig:hybrid_solution} illustrates the overall security architecture can be enabled by the proposed ASIC+eFPGA design. The framework spans the full lifecycle of the accelerator, beginning with design-time protections (S1 and S2). It then extends to runtime defenses, including side-channel shaping mechanisms that obscure observable hardware signatures (S3) as well as encryption and integrity verification of confidential model parameters during memory accesses (S4). Finally, the architecture incorporates deep pipeline monitoring to detect adversarial faults, metadata corruption, and aging-induced failures, while supporting long-term resilience through secure patching and rerouting around compromised control logic (S5). These mechanisms demonstrate how a bounded eFPGA fabric can serve as a reconfigurable security module within an ASIC-based LLM accelerator. Rather than relying on a static monolithic design, the hybrid eFPGA+ASIC architecture enables adaptive defenses that evolve with emerging hardware threats in edge AI systems.

\section{Evaluation: A Qualitative Comparison}

To compare mitigation strategies and architectural choices, Tables~\ref{tab:efpga_security_eval} and \ref{tab:arch_comparison} summarize their qualitative trade-offs, highlighting the design space of eFPGA-assisted security in ASIC-based LLM accelerators. Mechanisms such as selective IP redaction (S1) and device-bound configuration (S2) mainly enhance supply-chain trust and configuration integrity with modest overhead. In contrast, side-channel countermeasures (S3) and runtime monitoring (S5) provide active protection during execution, with moderate power and monitoring overheads. Integrating an eFPGA-based TEE (S4) offers strong confidentiality and integrity guarantees but incurs the highest performance and resource costs.

Table~\ref{tab:arch_comparison} positions the hybrid ASIC+eFPGA approach among accelerator architectures. ASIC-only designs offer maximum efficiency but limited post-silicon flexibility, while FPGA-based designs provide configurability at higher cost. The hybrid approach balances these by retaining ASIC efficiency and adding bounded reconfigurable security.
\begin{table}[t]
\centering
\caption{Design-space Evaluation of eFPGA-enabled Security Strategies.}
\label{tab:efpga_security_eval}
\scriptsize
\setlength{\tabcolsep}{2.5pt}
\begin{tabular}{lccccc}
\toprule
\textbf{Strategy} & \textbf{Security} & \textbf{Adaptability} & \textbf{HW Cost} & \textbf{Perf. Cost} & \textbf{Maturity} \\
\cmidrule[\heavyrulewidth](r){1-1} \cmidrule[\heavyrulewidth](r){2-2} \cmidrule[\heavyrulewidth](r){3-3} \cmidrule[\heavyrulewidth](r){4-4} \cmidrule[\heavyrulewidth](r){5-5} \cmidrule[\heavyrulewidth](r){6-6} 
S1: IP Redaction & $\bullet\bullet\bullet$ & $\bullet$ & $\bullet\bullet$ & $\bullet\bullet$ & Mature \\
S2: Encrypted Config & $\bullet\bullet\bullet$ & $\bullet\bullet$ & $\bullet$ & $\bullet$ & Mature \\
S3: SCR Randomization & $\bullet\bullet\bullet$ & $\bullet\bullet\bullet$ & $\bullet\bullet$ & $\bullet\bullet$ & Emerging \\
S4: eFPGA TEE & $\bullet\bullet\bullet$ & $\bullet\bullet\bullet$ & $\bullet\bullet\bullet$ & $\bullet\bullet\bullet$ & Research \\
S5: Monitors and Patching & $\bullet\bullet\bullet$ & $\bullet\bullet\bullet$ & $\bullet\bullet$ & $\bullet\bullet$ & Research \\

\bottomrule
\end{tabular}

\vspace{2pt}
\footnotesize
$\bullet$ Low \quad
$\bullet\bullet$ Medium \quad
$\bullet\bullet\bullet$ High
\end{table}

\begin{table}[t]
\centering
\caption{Comparison of Accelerators for Secure LLM Inference.}
\label{tab:arch_comparison}
\scriptsize
\setlength{\tabcolsep}{4pt}
\begin{tabular}{lccc}
\toprule
\textbf{Dimension} & \textbf{ASIC-only} & \textbf{FPGA-only} & \textbf{Hybrid ASIC + eFPGA} \\
\cmidrule[\heavyrulewidth](r){1-1} \cmidrule[\heavyrulewidth](r){2-2} \cmidrule[\heavyrulewidth](r){3-3} \cmidrule[\heavyrulewidth](r){4-4} 
Compute Efficiency & $\bullet\bullet\bullet$ & $\bullet$ & $\bullet\bullet$ \\
Security Flexibility & $\bullet$ & $\bullet\bullet\bullet$ & $\bullet\bullet$ \\
Post-deployment Updates & $\bullet$ & $\bullet\bullet\bullet$ & $\bullet\bullet$ \\
SCR Adaptability & $\bullet$ & $\bullet\bullet\bullet$ & $\bullet\bullet$ \\
Supply-chain Protection & $\bullet\bullet$ & $\bullet$ & $\bullet\bullet\bullet$ \\
Silicon Area Cost & $\bullet$ & $\bullet\bullet\bullet$ & $\bullet\bullet$ \\
\bottomrule
\end{tabular}

\vspace{2pt}
\footnotesize
$\bullet$ Low \quad
$\bullet\bullet$ Medium \quad
$\bullet\bullet\bullet$ High
\end{table}

\section{Future Trend and Conclusion}

Attacks on ASIC-based DNN accelerators highlight critical vulnerabilities across confidentiality, integrity, and availability, arising from threats such as hardware Trojans, physical side-channel leakage, and memory tampering or spoofing. These risks are further exacerbated in LLM accelerators, where high-value assets—including model weights and KV caches—coexist with deeply pipelined execution flows, making the system highly sensitive to even minor timing variations or fault injections that can propagate into significant inference errors. In this work, we advocate for the integration of a lightweight eFPGA fabric within ASIC-based accelerators as a means to introduce security agility. Such hybrid architectures enable selective logic redaction, adaptive side-channel mitigation, and enhanced memory protection mechanisms, while preserving the high performance and efficiency of ASIC implementations. Beyond these capabilities, eFPGA integration fundamentally shifts the security model from static to adaptive. As LLM accelerators continue to evolve toward distributed and large-scale architectures (e.g., wafer-scale or mesh-based systems), the hardware attack surface is expected to grow substantially. In this context, eFPGA-enabled capabilities, e.g., runtime monitoring, post-deployment patching, and adaptive security, become essential for sustaining long-term trust.

\bibliographystyle{IEEEtran}
\bibliography{refs}

\end{document}